\documentclass[twocolumn]{aastex631}

\usepackage{amsmath}
\usepackage{amssymb}
\usepackage{acronym}
\usepackage{hyperref}
\usepackage{color}
\usepackage{graphicx}
\usepackage{mathrsfs}
\usepackage{multirow}
\usepackage{comment}
\usepackage{ulem} 
\usepackage{tensind}
\tensordelimiter{?}
\tensorformat{lrc}

\usepackage{pifont} %

\newcommand{\ie}{\textit{i.e.}}
\newcommand{\eg}{\textit{e.g.}}
\newcommand{\be}{\begin{equation}}
\newcommand{\ee}{\end{equation}}
\newcommand{\bea}{\begin{eqnarray}}
\newcommand{\eea}{\end{eqnarray}}
\newcommand{\bel}{\begin{align}}
\newcommand{\eel}{\end{align}}

\def\i{{\rm i}}

\def\Msun{{\rm M_{\odot}}}
\def\Mo{{\Msun}}

\def\GMc2{{\rm G M_{\odot} c^{-2}}}

\def\knec{{\tt SNEC} }

\def\snec{{\tt SNEC} }
\def\skynet{{\tt SkyNet} }

\definecolor{cyan}{rgb}{0,0.9,0.9}
\definecolor{orange}{rgb}{0.9,0.5,0}
\definecolor{magenta}{rgb}{1,0,1}
\definecolor{purple}{rgb}{0.8,0.4,0.8}
\definecolor{gray}{rgb}{0.8242,0.8242,0.8242}

\newcommand{\newtxt}[1]{{#1}}

\begin{document}

\title{    Element formation in radiation-hydrodynamics simulations of kilonovae
}

\author[0009-0005-0976-7851]{Fabio Magistrelli}
\email{fabio.magistrelli@uni-jena.de}
\affiliation{Theoretisch-Physikalisches Institut, Friedrich-Schiller-Universit{\"a}t Jena, 07743, Jena, Germany}
   
\author[0000-0002-2334-0935]{Sebastiano Bernuzzi}
\email{sebastiano.bernuzzi@uni-jena.de}
\affiliation{Theoretisch-Physikalisches Institut, Friedrich-Schiller-Universit{\"a}t Jena, 07743, Jena, Germany}

\author[0000-0002-0936-8237]{Albino Perego}
\email{albino.perego@unitn.it}
\affiliation{Dipartimento di Fisica, Università di Trento, Via Sommarive 14, 38123 Trento, Italy}
\affiliation{INFN-TIFPA, Trento Institute for Fundamental Physics and Applications, via Sommarive 14, I-38123 Trento, Italy}

\author[0000-0001-6982-1008]{David Radice}
\email{david.radice@psu.edu}
\thanks{Alfred P.~Sloan fellow}
\affiliation{Institute for Gravitation \& the Cosmos, The Pennsylvania State University, University Park, PA 16802}
\affiliation{Department of Physics, The Pennsylvania State University, University Park, PA 16802}
\affiliation{Department of Astronomy \& Astrophysics, The Pennsylvania State University,University Park, PA 16802}

\date{\today}

\begin{abstract}
  Understanding the details of $r$-process nucleosynthesis in binary neutron star mergers (BNSM) ejecta is key to interpreting kilonova observations and identifying the role of BNSMs in the origin of heavy elements.
  We present a self-consistent two-dimensional, ray-by-ray radiation-hydrodynamic evolution of BNSM ejecta with an online nuclear network (NN) up to the days timescale.
  For the first time, an initial numerical-relativity ejecta profile composed of the dynamical component, spiral-wave and disk winds is evolved including detailed $r$-process reactions and nuclear heating effects. A simple model for the jet energy deposition is also included.
  Our simulation highlights that the common approach of relating in post-processing the final nucleosynthesis yields to the initial thermodynamic profile of the ejecta can lead to inaccurate predictions.
      Moreover, we find that neglecting the details of the radiation-hydrodynamic evolution of the ejecta in nuclear calculations can introduce deviations up to one order of magnitude in the final abundances of several elements, including very light and second $r$-process peak elements.
      The presence of a jet affects element production only in the innermost part of the polar ejecta, and it does not alter the global nucleosynthesis results.
  Overall, our analysis shows that employing an online NN improves the reliability of nucleosynthesis and kilonova light curves predictions.
\end{abstract}

\section{Introduction}
\label{sec:introduction}

Mass ejecta from binary neutron star mergers (BNSMs) are primary sites for rapid neutron capture ($r$-process) nucleosynthesis \citep{Eichler:1989ve}, see \eg~\citep{Cowan:2019pkx, Perego:2021dpw, Arcones:2022jer}.
The heavy, neutron-rich elements produced in these environments undergo radioactive decays, powering an electromagnetic (EM) transient known as a kilonova. Moreover, some of these mergers can produce short gamma-ray bursts (sGRBs),
offering further insights into their astrophysical properties. The unambiguous detection of gravitational waves (GW170817) and EM counterparts (kilonova AT2017gfo and GRB 170817A) from a BNSM in August 2017 has confirmed theoretical predictions and triggered intense work on the subject~\citep{LIGOScientific:2017vwq, GBM:2017lvd, LIGOScientific:2017zic, Coulter:2017wya, Tanvir:2017pws, Savchenko:2017ffs}.

Numerical simulations are the main tool to explore the physics of BNSMs and identify the mechanisms underlying the associated gravitational and EM emissions \citep{Rosswog:2013kqa, Roberts:2016igt, Nedora:2020hxc, Radice:2021jtw, Zappa:2022rpd, Combi:2022nhg, Combi:2023yav, Schianchi:2023uky, Radice:2023zlw, Kiuchi:2023obe, Musolino:2024xqy}.
Simulations are used to carry out extensive investigations on $r$-process nucleosynthesis and its relationship with kilonovae.
Lanthanides and actinides production is consistently found in low electron fraction ejected material ($Y_e \lesssim 0.2$) \citep{Korobkin:2012uy, Lippuner:2015gwa, Perego:2021dpw}. Low values of $Y_e$ can be found in the equatorial component of dynamical ejecta \citep{Wanajo:2014wha, Radice:2016dwd, Kiuchi:2022nin} and in the late disk wind, \eg~\citep{Beloborodov:2002af, Siegel:2017nub, Sprouse:2023cdm, Kiuchi:2023obe}. The high opacity associated with these elements commonly links the outer regions of the (low-latitude) ejecta with the red component of the kilonova light curves \citep{Metzger:2014ila, Perego:2017wtu}.
However, small expansion timescales can inhibit neutron captures and thus produce a blue-UV precursor powered by decays of free neutrons \citep{Metzger:2014yda, Radice:2018ghv, Combi:2022nhg}.
Higher values of $Y_e \gtrsim 0.3$ associated with most of the disk and the majority of polar ejecta prevent strong $r$-processes, thus leading to low opacity matter contributing to the blue part of the kilonova spectra \citep{Metzger:2014ila, Kasen:2017sxr, Tanaka:2017qxj, Curtis:2023zfo}. In the polar regions, this is enhanced by interactions with neutrinos emitted by the central remnant until its possible collapse to a black hole \citep{Radice:2023xxn}.
The passage of a relativistic jet can in principle affect the nucleosynthesis by injecting extra energy in the system and thus reigniting suppressed reactions \citep{Janiuk:2014tga}.
Understanding the details of the nuclear evolution of the ejecta and how this affects its dynamics is crucial for accurately interpreting observational data and assessing the role of BNSMs in explaining the origin of $r$-process elements in the universe.

One of the main open challenges towards the accurate prediction of $r$-process nucleosynthesis in BNSMs ejecta is capturing the dependence of the nuclear composition outcomes on the initial thermodynamic profile and detailed hydrodynamic evolution of the ejecta.
Traditional approaches assume homologously expanding ejecta and neglect radiation transport, hydrodynamics and ejecta self-interaction, \eg~\citep{Korobkin:2012uy, Radice:2016dwd, Rosswog:2016dhy, Perego:2020evn, Curtis:2023zfo}, although some recent efforts have considered detailed radiation transport, \eg~\citep{Collins:2022ocl, Shingles:2023kua}.
Alternatively, nuclear networks are employed in a post-processing step on hydrodynamics profiles to obtain yields and heating rates \citep{Just:2014fka, Martin:2015hxa, Roberts:2016igt}. The latter are sometimes fitted as time-domain functions and re-used in hydrodynamical simulations to improve ejecta evolutions, \eg~\citep{Rosswog:2013kqa, Wu:2021ibi, Ricigliano:2023svx}.

In this work, we present, for the first time, a self-consistent simulation of $r$-process nucleosynthesis for a fiducial BNSM ejecta using a 2-dimensional, ray-by-ray radiation hydrodynamics code coupled with an online nuclear network (NN). In Section \ref{sec:method}, we present our model and discuss how the NN is coupled with the radiation-hydrodynamic equations to update the ejecta composition and calculate the associated nuclear power in each hydrodynamic step. We also describe our implementation of thermalization processes and of an extra energy source modeling a sGRB. Section \ref{sec:results} presents the nucleosynthesis and kilonova light curves predictions for the ejecta profiles we extract from a fiducial BNSM numerical relativity simulation.
In Section \ref{sec:conclusion}, we conclude discussing the implications of our findings and emphasizing the importance of implementing an online NN for self-consistent predictions of $r$-process nucleosynthesis patterns.

\section{Method}
\label{sec:method}

\subsection{Numerical-relativity profiles}
\label{subsec:init-cond}

Our simulation is initialized with initial ejecta profiles extracted from ab-initio numerical relativity simulations that include microphysics, M0 neutrino transport\footnote{See \cite{Zappa:2022rpd} for the importance of including neutrino heating effects.} and magnetic-field induced turbulence \citep{Radice:2018pdn, Perego:2019adq, Bernuzzi:2020tgt, Nedora:2020hxc}.
The considered binary has mass ratio  $q=1.43$ and produces a short-lived remnant that collapses to a black hole (BH) at about $10$~ms postmerger. The simulation is run with the LS220 equation of state (EoS) and with a typical resolution of $185$~m \newtxt{up to $28.5$~ms after merger} \citep{Nedora:2020hxc}. Throughout the simulation, the unbound material is identified via the Bernoulli criterion and collected at an extraction radius of $R_\text{ext} \simeq 295$~km. The ejecta consists of the dynamical component of mass ${\sim} 1.24 \times 10^{-2}~\Mo$ and a short spiral-wave wind (${\sim} 0.80 \times 10^{-2}~\Mo$). In order to prepare the subsequent long-term evolution, the dependence of the ejecta properties on the azimuthal coordinate is integrated out, while the polar dependence is accounted for by discretizing the polar angle in $51$ angular sections. The time-dependent ejecta profile is mapped to a Lagrangian profile by positioning the latest shell at $R_\text{ext}$ and progressively layering previously ejected shells on top as prescribed by the constraint $m(r) = 4\pi \int_{R_\text{ext}}^{r} \rho(r) \, r^2 dr$, with $\rho$ mass density and $r$ position of the radius including the mass $m$ \citep{Wu:2021ibi}.
An analytical disk wind profile of mass ${\sim} 2.75 \times 10^{-2}~\Mo$ is constructed from a similar simulation but performed on an equal mass binary with the M1 neutrino transport \citep{Radice:2021jtw}, which better describes the optically thick regime \citep{Zappa:2022rpd}. The simulation of the $q=1$ binary is run with the SFHo EoS for ${\sim} 270$~ms postmerger\footnote{
	We take the results of this simulation, where the BH collapse occurs around $t \simeq 10 \textrm{ ms}$ postmerger, as a representative history of disk ejecta.
	The general properties of disks and disk ejecta are found to be relatively stable against variations in the properties of the compact binary or numerical prescriptions (\eg~EoS, effective viscosity, neutrino treatment), as long as the BH collapse occurs in a timescale of $\mathcal{O}(10) \textrm{ ms}$ \citep{Fernandez:2013tya, Just:2021cls, Just:2023wtj, Kiuchi:2022nin, Kiuchi:2023obe, Camilletti:2024otr}.}.
The disk profile is constructed by interpolating the spherically averaged density, temperature, and electron fraction of the unbound matter over time and using second-order Pad\'e approximants. We assume a constant velocity and entropy, and rescale the density evolution to get a total ejection of $40\%$ of the total disk mass \newtxt{(of the LS220 simulation)} assuming a $\propto\sin^2\!\theta$ angular distribution~\citep{Perego:2017wtu, Fernandez:2018kax}. The complete initial profiles of two selected angular sections are shown in the two bottom panels of Fig.~\ref{fig:time_mass_comp}. 
\subsection{Ray-by-ray Radiation-Hydrodynamics}
\label{subsec:rad-hydro}

The ejecta profile is evolved with the system of Lagrangian radiation-hydrodynamic equations as implemented in \citet{Morozova:2015bla,Wu:2021ibi}. In particular, the energy equation in spherical symmetry reads
\begin{equation}
  \label{eq:energy_cons}
  \frac{\partial \epsilon}{\partial t} = \frac{P}{\rho} \, \frac{\partial \ln\rho}{\partial t} - 4\pi r^2 Q \frac{\partial v}{\partial m} - \frac{\partial L}{\partial m} + \dot{\epsilon}_\text{nucl} \,,
\end{equation}
with $m$ mass coordinate, $\epsilon$ specific internal energy, $t$ time, $P$ pressure, $v$ (radial) velocity, $L$ luminosity and $Q$ von Neumann-Richtmyer artificial viscosity \citep{VonNeumann:1950}. The specific energy deposition $\dot{\epsilon}_\text{nucl}$ accounts for the local production and partial thermalization of the energy released by all possible nuclear reactions occurring in the expanding material. To calculate the luminosity, we employ the same analytic, time-independent opacity introduced in \cite{Wu:2021ibi}.

The set of hydrodynamic equations is closed by the EoS. In the high-temperature regime, we implement the tabulated Helmholtz EoS introduced in \cite{Timmes:2000a} and also used in \cite{Lippuner:2017tyn}. For lower temperatures, where the contribution of positrons is negligible, we switch to the Paczynski EoS introduced in \cite{Paczynski:1986px} and utilized in \cite{Morozova:2015bla, Wu:2021ibi}. This analytical EoS describes a mixture of ideal gases, namely a Boltzmann gas of non-degenerate, non-relativistic ions, an ideal gas of arbitrarily degenerate and relativistic electrons, and a photon ideal gas. The inclusion of an online nuclear network, as described in Sec.~\ref{subsec:nnCoupling}, improves the realism of these EoS, as the mean molecular weight and the electron fraction, which enter the EoS, are self-consistently calculated on the fly.

To account for the angular dependency of the ejecta properties, we employ the spherically symmetric hydrodynamic equations in a ray-by-ray fashion. For each angular section, we first map the profile into an effective 1D problem by multiplying the total mass by the scaling factor $\lambda_\theta = 4\pi / \Delta\Omega$, where $\Delta\Omega \simeq 2\pi \sin{\theta} \, d\theta$ represents the solid angle included in the angular section. This ensures that all the intensive quantities (including density) remain fixed.
Then, we independently evolve the different angular sections by discretizing the 1D hydrodynamic equations over $n_\text{sh}$ spherical fluid elements (mass shells). \newtxt{We ran tests for some selected sections with grid resolutions $n_\text{sh} = 300, 600, 1000$. We found $n_\text{sh} = 600$ sufficiently high to resolve the details of the nucleosynthesis and decided to use this resolution for the complete simulation.} Note that non-radial flows of matter and radiation are neglected, \newtxt{as well as higher dimensional fluid instabilities (\eg~Kelvin-Helmholtz), convection and the possibility of shells surpassing each other. This could in principle lead to over or underestimating the pressure work exchanged between the radial shells and introduce artificial shocks in the system.} Finally, we map the problem back to the axisymmetric scenario by keeping the intensive quantities unchanged, rescaling the extensive ones by the scaling factor $1/\lambda_\theta$ and combining the results. In particular, we calculate the global mass fractions and abundances with a mass weighted average over all the mass shells and angular sections.
Kilonova light curves are recombined accounting for the angle of view as in \cite{Martin:2015hxa, Perego:2017wtu}. 

\subsection{Nuclear Network Coupling}
\label{subsec:nnCoupling}

We calculate the nuclear composition and specific energy deposition, $\dot{\epsilon}_\text{nucl}$, in a self-consistent way with an online implementation of the nuclear network (NN) \skynet \citep{Lippuner:2017tyn}.
The NN includes 7836 isotopes up to~$^{337}$Cn and uses the JINA REACLIB \citep{Cyburt:2010a} and the same setup as in \cite{Lippuner:2015gwa, Perego:2020evn}. 
The simulation is started by initializing matter composition from the initial temperature $T_0 (m)$ and density $\rho_0 (m)$ assuming nuclear statistical equilibrium (NSE). We thus impose, to each isotope considered in the NN, the abundance \citep{Cowan:2019pkx, Perego:2021dpw}
\begin{equation}
	\label{eq:NSEcomp}
	\begin{split}
		Y_{i,0} = \frac{n_{i,0}}{n_{b,0}} =
		Y_{p,0}^{Z_i} \, Y_{n,0}^{(A_i - Z_i)} \, \frac{G_i(T) \, A_i^{3/2}}{2^{A_i}} \left(\frac{\rho}{m_b}\right)^{\!\! A_i - 1} \\
		\times
		\left(\frac{2\pi \hbar^2}{m_b k_\text{B}T}\right)^{\!\! 3(A_i-1)/2} e^{\text{BE}_i/k_\text{B}T} \,,
	\end{split}
\end{equation}
with $m_b$ and $n_b$ baryon mass and number density, $A_i$ and $Z_i$ mass and atomic numbers of the $i$-th isotope and $n_i$ and $\text{BE}_i$ its number density and total binding energy\footnote{
						\newtxt{Several mass shells have $T < 5$~GK at the end of the numerical relativity simulation (see Fig.~\ref{fig:time_mass_comp}). In the case of low entropy tidal ejecta, their initial composition should be close to the one set by cold NSE conditions inside isolated neutron stars. In all the other cases, we expect matter to reach hot NSE conditions inside our extraction radius, before matter expansion causes an NSE freeze-out. We plan to attest the impact of the early out-of-NSE evolution on our results by developing and comparing alternative initialization procedures for cold fluid elements in a future work.}}.
	Here, $G_i(T)$ is the internal partition function of the nucleus $i$ at temperature $T$. The two conditions $\sum_i A_i Y_i = 1$ and $\sum_i Z_i Y_i = Y_e$, with $Y_e$ electron fraction, constrain the NSE composition, which thus depends on $\rho$, $T$ and $Y_e$.
See \cite{Lippuner:2017tyn} for details about \skynet\!\!'s implementation of Eq.~\eqref{eq:NSEcomp}.
The isotopic mass fractions can be calculated from the element abundances via $X_i = A_i Y_i$.

\newtxt{At the initial time $t_0$ of each hydrodynamic step $\Delta t$, an independent instance of \skynet is called in each mass shell.
The NN evolves the composition with its own time sub-steps until $t_0 + \Delta t$.
The (non-thermalized) power released by nuclear reactions is averaged over $\Delta t$ to get the total nuclear power that will be then thermalized into $\dot{\epsilon}_\text{nucl}$ as described below.
During the sub-steps, the temperature is constant (no self-heating) and fixed by $T(t_0)$, and it will only be updated by the hydrodynamic step. The density evolution is prescribed with a log-interpolation between $\rho(t_0)$ and $\rho(t_0 + \Delta t)$. The latter is calculated from the density and radial velocity profiles at $t=t_0$. Note that in our setup mass shells cannot mix.}

To calculate the specific energy deposition in Eq.~\eqref{eq:energy_cons}, we independently thermalize the different contributions to $\dot{\epsilon}_\text{nucl} = \dot{\epsilon}_\text{nucl}^\gamma + \dot{\epsilon}_\text{nucl}^\alpha + \dot{\epsilon}_\text{nucl}^\beta + \dot{\epsilon}_\text{nucl}^\text{ oth}$ coming from $\gamma$ rays, $\alpha$ particles, electrons, and other nuclear reactions products.
We compute the fraction of energy thermalized by the emission of $\gamma$-rays as
\begin{equation}
	\dot{\epsilon}_\text{nucl}^\gamma(t) = \sum_j f_j^\gamma (t) \frac{\langle E_j^\gamma \rangle Y_j(t)}{\tau_j \, m_p} \,,
\end{equation}
with $j$ isotope index and $m_p$ proton mass. The average lifetimes $\tau_j$ and the mean energy $\langle E_j^\gamma \rangle$ released by each of these nuclei in $\gamma$-rays via a radioactive decay are taken from the ENDF/B-VIII.0 database\footnote{
	\url{https://www-nds.iaea.org/exfor/endf.htm}}
\citep{Brown:2018jhj}.
The thermalization factor $f_j^\gamma (t)$ is calculated, as in \cite{Hotokezaka:2019uwo, Combi:2022nhg}, starting from the detailed composition of the ejecta and the same effective opacity tables from the NIST-XCOM catalogue\footnote{
	\url{https://www.nist.gov/pml/xcom-photon-cross-sections-database}}
\citep{Berger:2010} used in \cite{Barnes:2016umi}.
For the energy injected in electrons and $\alpha$ particles and their thermalization factors, we use the analytic expressions of \cite{Kasen:2018drm}. Neutrinos from $\beta$-decays do not deposit energy into the system, while fission fragments and daughter nuclei tend to thermalize very efficiently \citep{Barnes:2016umi}. We approximate this behavior by assuming that half of the remaining (non-thermalized) heating rate from \skynet is deposited into the ejecta. 

\subsection{Jet Energy Deposition}
\label{subsec:sGRB_model}

We consider a jet model following the ``thermal bomb'' prescription of \cite{Morozova:2015bla}. Essentially, an extra energy term is added to the RHS of Eq.~\eqref{eq:energy_cons} for the innermost shells of the ejecta during a chosen time interval. The thermal bomb parameters are fixed assuming that the released isotropic energy from the sGRB is proportional to the kinetic energy of the structured jet described by \cite{Ghirlanda:2018uyx}, $E_\text{iso} (\theta) = E_0 / [1 + (\theta/\theta_j)^{5.5}]$, with $\theta_j$ jet's opening angle. The total energy $E_j$ released by the jet is related to $E_\text{iso}$ by $E_\text{iso} = (4\pi / \Delta\Omega_j) E_j$, where $\Delta\Omega_j = 2\pi (1 - \cos\theta_j)$ is the solid angle covered by the jet. We fix $\theta_j = 15$ degrees, within the range also explored in \cite{Hamidani:2019qyx} for the sGRB associated with GW170817. We simulate a jet with the same duration as the observed sGRB from \cite{LIGOScientific:2017zic}, $\Delta t_j = 100 \textrm{ ms}$.
We launch the jet at $t = 200 \textrm{ ms}$ and fix the parameter $E_0 = 10^{51} \textrm{ erg}$, to be compatible with the isotropic luminosity range given by \cite{Hamidani:2019qyx}. Their Figure~9 shows that, in the case of GRB 170817A, choosing the isotropic luminosity to be $L_\text{iso} \sim 10^{52} \textrm{ erg/s}$ constrains the jet to be launched no more than $300 \textrm{ ms}$ after merger.

\section{Results}
\label{sec:results}

\subsection{Fluid Elements Composition Evolution}
\label{subsec:comp_evol}

\begin{figure*}[t]
	\centering 
	\includegraphics[width=\textwidth]{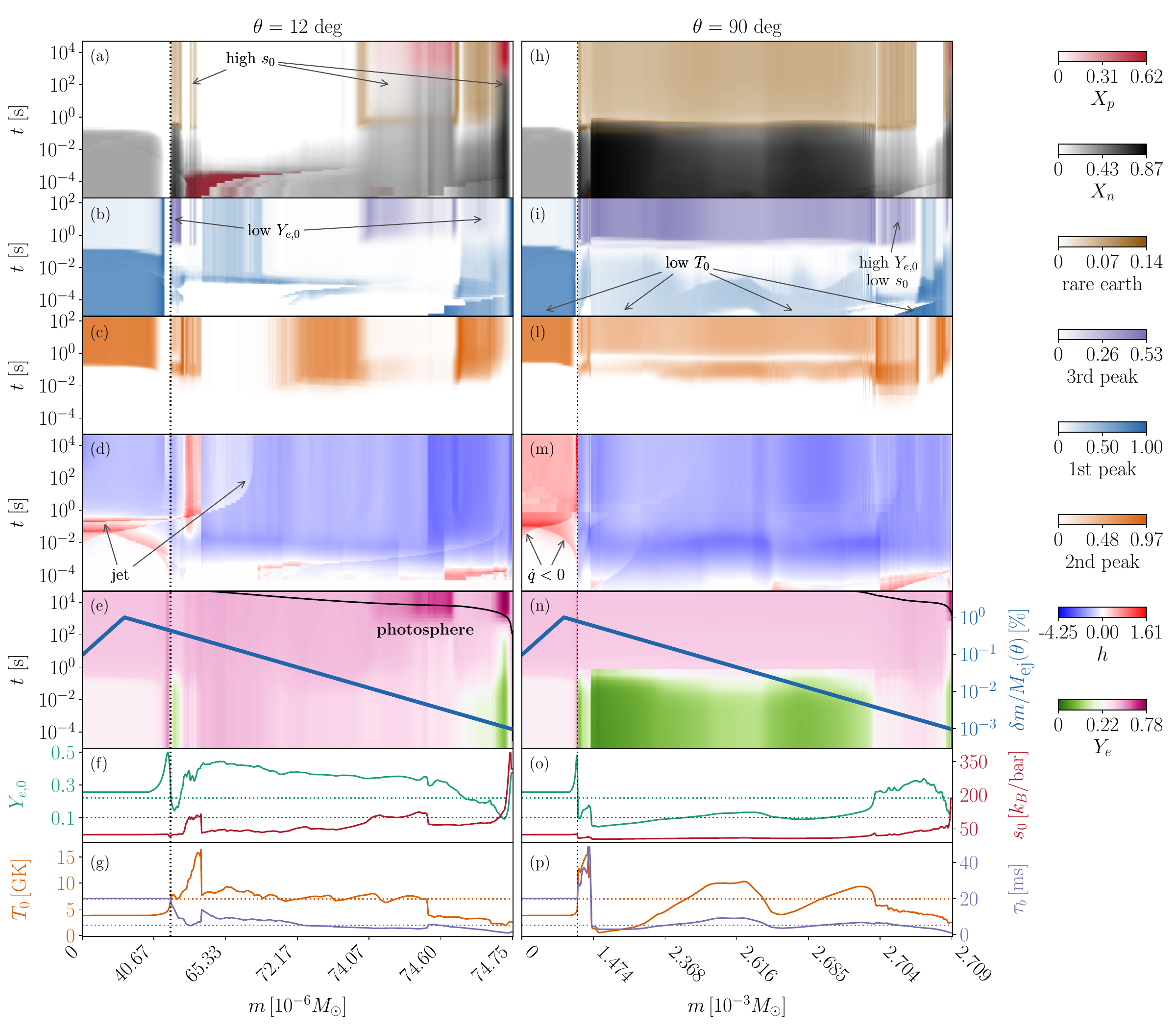}
	\caption{Evolution of the ejecta composition for every mass shell of the angular sections $\theta \simeq 12,90$ degrees (first and second column respectively). The x-axis indicates the Lagrangian mass coordinate\newtxt{, linear in the shell index. The vertical black, dotted lines indicate the point where the analytical disk wind is attached to the dynamical ejecta profile.}
		The color maps in panels ($a$-$c$) and ($h$-$l$) represent the mass fractions of neutrons (in gray), protons (red), first (blue), second (orange), and third (purple) r-process peak elements, and rare earths (brown). Panels ($d$, $m$) and ($e$, $n$) show the evolution of the parameter 		$h(t)$ defined in \eqref{eq:homol_par} (blue to red) and of the electron fraction (green to pink), respectively.
		In ($e$) and ($n$) we also plot 		\newtxt{in blue the amount of mass $\delta m$ included in each shell, rescaled by the total mass of the angular section $M_\text{ej}(\theta)$,}
		and \newtxt{in black} the position of the photosphere. 		Subplots ($f$-$g$) and ($o$-$p$) show the initial values of $Y_e, s, T, \tau$ as a function of the included mass. 		The horizontal dotted lines indicate the limit values of $Y_{e,0} = 0.22$, $s_0 = 100 \, k_\text{B}/\text{baryon}$, $T_0 = 7 \text{ GK}$, and $\tau_b = 5 \textrm{ ms}$.}
	\label{fig:time_mass_comp}
\end{figure*}

Our simulation allows to self-consistently monitor the nuclear composition of the matter during its expansion.
Figure~\ref{fig:time_mass_comp} shows the time evolution of the mass fraction of protons, neutrons, rare earths and $r$-process peak elements (panels ($a$-$c$) and ($h$-$l$)), of the parameter $h(t)$ (see below, panels ($d$) and ($m$)) and of the electron fraction (panels ($e$) and ($n$)), together with the initial thermodynamic conditions of the ejecta (panels ($f$,$g$) and ($o$,$p$)), as a function of the Lagrangian mass coordinate.
The left and right panels refer to two different angular sections, $\theta = 12,\, 90$ degrees.
Mass shells with, respectively, $m \lesssim 50 \times 10^{-6}~\Mo$ and $m \lesssim 1.4 \times 10^{-3}~\Mo$ correspond to the wind, while larger values to the dynamical ejecta. In the latter, the entropy slightly increases towards the outer shells, where a high entropy, low mass tail is observed at all angles.
Material with low initial electron fractions, $Y_{e,0} < 0.22$, is found in the outer regions of the ejecta and, for $m \sim 2.5 \times 10^{-3}~\Mo$, around the equatorial plane. During the evolution, the electron fraction rises to $Y_e > 0.22$ for almost every mass shell at around $t \sim 1 \textrm{ s}$ due to the $\beta$-decays following the neutron captures and neutron freeze-out. 

The color maps in panels ($d$) and ($m$) of Fig.~\ref{fig:time_mass_comp} show the evolution of the parameter \begin{equation}
	\label{eq:homol_par}
	h(t) \equiv \log_{10} \left( \frac{\rho(t)}{\rho_\skynet(t\,|\,\rho_0,\tau_b)} \right) \,.
\end{equation}
The density evolution for the analytical expansion $\rho_\skynet$ is parametrized by the initial density $\rho_0$ and expansion timescale $\tau_b$ as defined in Eq. (1) of \cite{Lippuner:2015gwa}.
\newtxt{Here, $\tau_b$ is calculated from the expansion timescale measured at the extraction radius as prescribed in \citep{Radice:2016dwd}.}
The same homologous expansion is used by \cite{Perego:2020evn, Wu:2021ibi} for the time-domain fitting of the heating rates.
Positive (negative) values of $h$ indicate regions of the $t-M$ plane where the mass shells are characterized by higher (lower) densities than the one predicted by the simple analytical model. Shells where $h$ is constant in time are expanding homologously. This happens for most of the shells in our simulation for $t \gtrsim 100 \textrm{ s}$, in agreement with the results of \cite{Rosswog:2013kqa}.
If $h \neq 0$, the evolution happens on a curve $\rho_\skynet(t\,|\,\rho'_0,\tau'_b)$ different from the one associated with the values of $\rho_0, \tau_b$ of the considered mass shell.
We identify a deviation of approximately two to three orders of magnitude between our numerical solution and the specific analytical expansion $\rho_\skynet(t\,|\,\rho_0,\tau_b)$ \newtxt{in all the significative timescales}.
As further explained below, such a different expansion \newtxt{in the early evolution ($t \lesssim 1$~s, when the production of heavy nuclei is still ongoing)} can lead to inconsistent nucleosynthesis predictions.
\newtxt{In Appendix \ref{app:hydro_on_NN} we discuss an example of how a full evolution of a fluid element can differ from the one obtained by an independent NN assuming $\rho_\skynet(t\,|\,\rho_0,\tau_b)$.}
\newtxt{A comparison with the results from a run using the fitted heating rates of \cite{Wu:2021ibi} instead of the data from the coupled NN reveals that the disagreement between the predicted density evolution and the simulated one builds up mostly because of hydrodynamic effects. The relative differences in the nuclear heating are not big enough to impact the hydrodynamic evolution while competing with the pressure work in Eq.~\eqref{eq:energy_cons} (see Appendix \ref{app:NN_on_hydro}).}

At $t \sim 100 \textrm{ ms}$, a sudden increase of $h$ is observed in the disk ejecta. \newtxt{The analysis of the released nuclear energy reveals a correspondent} absorption of energy from the fluid to the NN ($\dot{\epsilon}_\text{nucl} < 0$). \newtxt{In the same region, the abrupt change in the $h$ parameter reveals ongoing shocks. A series of shock-induced photodissociation processes could explain the negative heating rates, in analogy with supernovae simulations, see \eg~\citep{Janka:2012wk, Burrows:2012ew}.} \newtxt{Such kind of} effect is genuinely due to the coupling between the NN and hydrodynamics, and it cannot be reproduced by offline analyses.
In the polar regions, a further compression at $t \sim 200 \textrm{ ms}$, followed by a more rapid expansion, is caused by the extra energy deposited by the jet.

Panels ($a$-$c$) and ($h$-$l$) of Fig.~\ref{fig:time_mass_comp} show the evolution of the cumulative mass fractions of groups of selected isotopes.
Consistently throughout the ejecta, shells with $Y_{e,0} \lesssim 0.22$ produce first peak elements (embedded in a sea of free neutrons) within $t \lesssim 10^{-4} \textrm{ s}$. On a time scale of a second, these nuclei are converted to second and third peak elements and a small fraction of rare earths.
At $T_0 \gtrsim 7 \text{ GK}$, most of the ejecta is initially composed of free neutrons and protons (their ratio depending on the initial electron fraction) and light elements, which will act as seeds for $r$-process nucleosynthesis.
In regions where $T_0 \lesssim 7 \text{ GK}$, the initial temperature is low enough in certain mass shells (see the arrows in panel (i)) to make the binding energy term in Eq.~\eqref{eq:NSEcomp} strongly affect the NSE composition. Hence, a significant fraction of first peak elements is already formed at $t = 0$ and can be directly used as r-process seeds. This can ease the heavy elements production and, in some cases, partially induce strong $r$-processes even for $Y_e \gtrsim 0.22$ (see for example the high $Y_{e,0}$, low $s_0$ third peak production in the $\theta = 90$ degrees case).

High enough values of the initial entropy ($s_0 \gtrsim 100 \, k_\text{B} / \text{baryon}$) can lead to the production of rare earths, lanthanides and actinides even for intermediate values of the initial electron fraction, $0.22 \lesssim Y_e \lesssim 0.35$. This outcome is compatible with an $\alpha$-rich freeze-out, characterized by a large neutrons to seed ratio \citep{Just:2014fka, Cowan:2019pkx}.
However, not all shells with high initial entropy are sites of strong $r$-process nucleosynthesis. This is clearly shown, for example, by the $\theta = 12$ degree angular section in the region $74.07 \lesssim m/(10^{-6}~\Mo) \lesssim 74.60$, \newtxt{see panels ($a$) and ($b$)}.
When strong $r$-processes are activated, free neutrons are efficiently consumed, reducing the final abundance of free protons. \newtxt{Starting from $m \sim 74.07 \times 10^{-6} M_{\odot}$ and moving outward in the ejecta,} while the initial $T_0$, $s_0$ and $\tau_b$ profiles are almost constant \newtxt{(panels ($f$) and ($g$)), strong $r$-processes get progressively suppressed. The suppression is smooth even given the jump in the initial conditions around $m \sim 74.60 \times 10^{-6} M_{\odot}$, which also does not change the second peak production in panel ($c$). This discussion gives an example of} how only the detailed evolution of the local thermo- and hydrodynamics conditions determine with accuracy the subsequent nucleosynthesis yields.

For low entropies, $s_0 \lesssim 100 \, k_\text{B} / \text{baryon}$, and intermediate values of the initial electron fractions, $0.22 \lesssim Y_{e,0} \lesssim 0.35$, the fewer free neutrons are rapidly captured by seed nuclei, and only second peak elements are produced at the end of the nucleosynthesis. Note again the exception represented by the shells at around $m \gtrsim 2.704 \times 10^{-3}~\Mo$ in the equatorial section.
At the high $Y_{e,0} \gtrsim 0.35$ reached in the polar section, weak $r$-processes occur at low entropy and only first peak elements can be produced.

Extremely high values of the initial entropy, $s_0 \gtrsim 200 \, k_\text{B} / \text{baryon}$, can be reached in the outermost shells of the ejecta.
Despite the large neutrons to seed ratio, neutron capture is not effective in the fast ejecta tail ($v \gtrsim 0.6$).
Nucleosynthesis is therefore hindered. In this rarefied environment, most of the neutrons remain free after the $r$-process freeze-out and start $\beta$-decaying on a timescale of $t \sim 10$ minutes, while crossing the photosphere, whose position is tracked in panels ($e$) and ($n$). This could power a UV/blue kilonova precursor on the hours timescales, as discussed in Sec.~\ref{subsec:light_curves}.
After $10^4 \text{ s}$, no heavy element is produced, and free protons dominate the final matter composition.
This is reflected in the high final electron fractions of these shells.
If the strong rise in entropy is combined with an initial electron fraction $Y_{e,0} \gtrsim 0.35$, all neutrons are initially bound in stable first peak elements. In these outer regions, the composition remains essentially frozen. We can thus predict a negligible nuclear contribution to the kilonova light curves from these shells throughout the ejecta evolution, regardless of the photosphere position.

The qualitative changes in the nucleosynthesis patterns caused by the jet are negligible. The only quantitative effect observed on the mass fractions depicted in Fig.~\ref{fig:time_mass_comp} is a very slight delay in the production of second $r$-process peak elements in the innermost shells of the polar sections. No significant effects arise in the regions near the equatorial plane, as the jet energy is negligible at low latitudes, and we do not account for the coupling between different angular sections.

In summary, our results indicate that it is not possible to predict the final nucleosynthesis yields based on the initial thermodynamic conditions of the ejecta solely.

\subsection{Global Nucleosynthesis Yields}
\label{subsec:nucl_yields}

\begin{figure}	\centering 
	\includegraphics[width=0.48\textwidth]{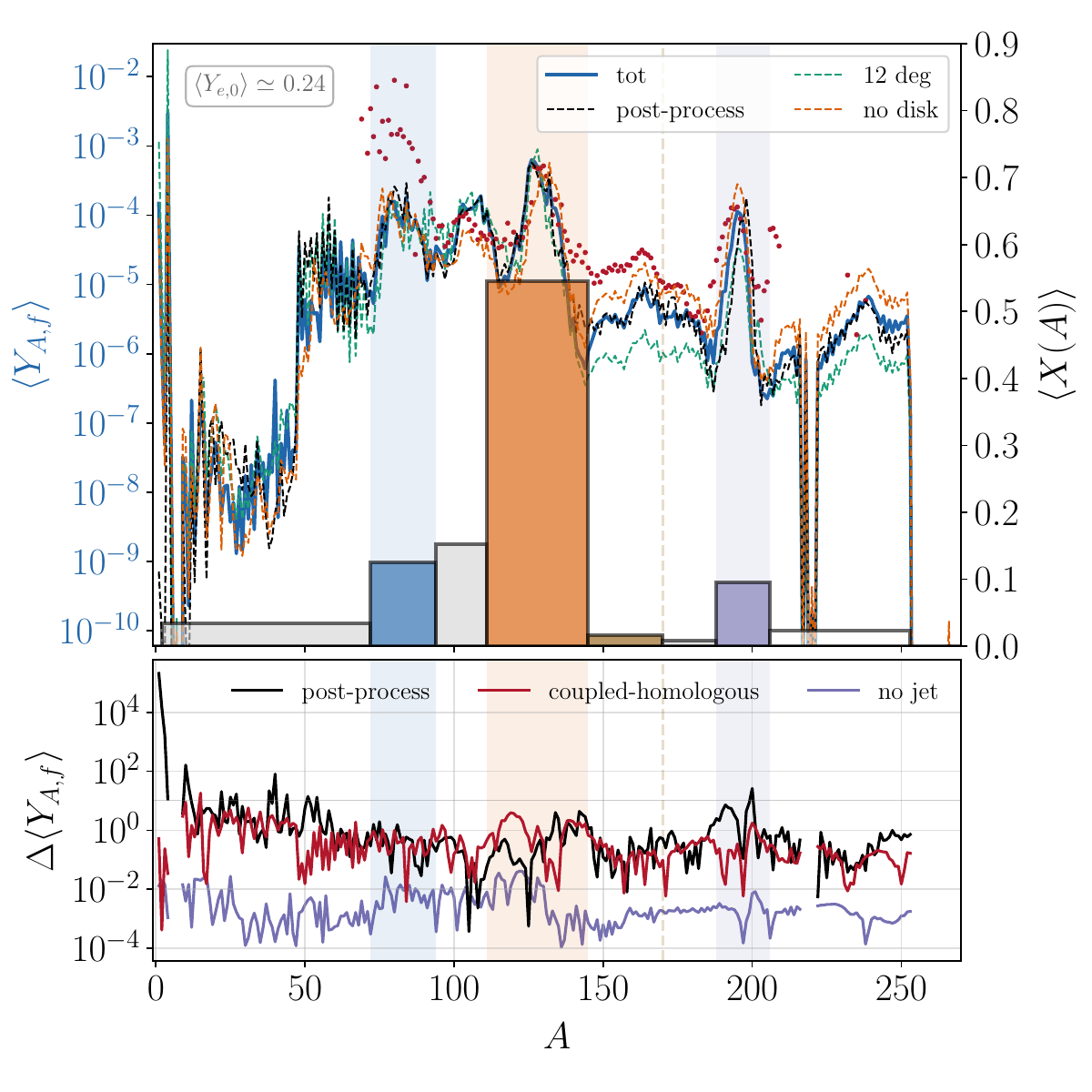}
	\caption{Global $r$-process nucleosynthesis yields at $t \simeq 5 \times 10^4 \textrm{ s}$. Top panel: in blue \newtxt{and black}, the final global abundances \newtxt{$\langle Y_{A,f} \rangle$ as a function of the mass number $A$, obtained respectively from our \knec simulation and with a post-process analysis similar to \cite{Perego:2020evn}.} 	In orange and green, the results obtained \newtxt{from our \knec simulation excluding the disk ejecta or including only the matter emitted at $\theta=12$ degrees (at all times)}. Red dots represent the Solar $r$-abundances \citep{Prantzos:2020a} scaled to match the average value of the second peak. The histogram shows, for the complete simulation, the global cumulative mass fractions $\langle X(A) \rangle$ of selected groups of elements; in particular, first (blue), second (orange) and third (purple) $r$-process peaks and rare-earths (brown). We report the initial value of the average electron fraction of the ejecta in the top left corner of the plot. 	Bottom panel: in red, relative difference \newtxt{against the model with prescribed homologous expansion}, see Eq.~\eqref{eq:rel_diff_abs}; in black and purple, analogous comparisons between our global results and the post-process analysis, and between the results of the complete simulation with and without the jet.
	\newtxt{The differences are only shown for mass numbers $\hat{A}$ such that $\langle Y_{\hat{A},f} \rangle > 10^{-10}$.}}
	\label{fig:final_abs}
\end{figure}

We calculate the global abundances $\langle Y_i \rangle$ as mass-weighted averages over all the mass shells of all the angular sections. In Fig.~\ref{fig:final_abs}, we show the matter composition at $t \simeq 5 \times 10^4 \textrm{ s}$. Except for small modifications due to long lived $\alpha$-decaying isotopes \newtxt{with $A \gtrsim 220$}, this is a good representation of the final $r$-process nucleosynthesis yields. All the $r$-process peaks and in part rare earths are produced. This is expected, as matter is mostly ejected along the equatorial plane, where the neutron-rich dynamical ejecta keeps the global, initial average electron fraction at $\langle Y_{e,0} \rangle \simeq 0.24$. \newtxt{The production of heavy elements is partially hindered in the} polar sections, \newtxt{consistently with their higher $Y_e$ and the picture outlined in Fig.~\ref{fig:time_mass_comp}}.
By comparing the results of our complete model with the ones obtained by excluding the late disk wind, we establish that the \newtxt{relative} production of heavy elements is already saturated in the dynamical ejecta. 
\newtxt{Most of the lighter elements} with \newtxt{$A \lesssim 130$} (in particular \newtxt{elements between the} first and second peaks and isotopes of the iron group) are \newtxt{relatively more} produced by the disk wind.
These results are in agreement with \cite{Martin:2015hxa, Cowan:2019pkx, Curtis:2022hjy, Chiesa:2023jno}. 
\begin{figure*}[t]
	\centering 
	\includegraphics[width=0.98\textwidth]{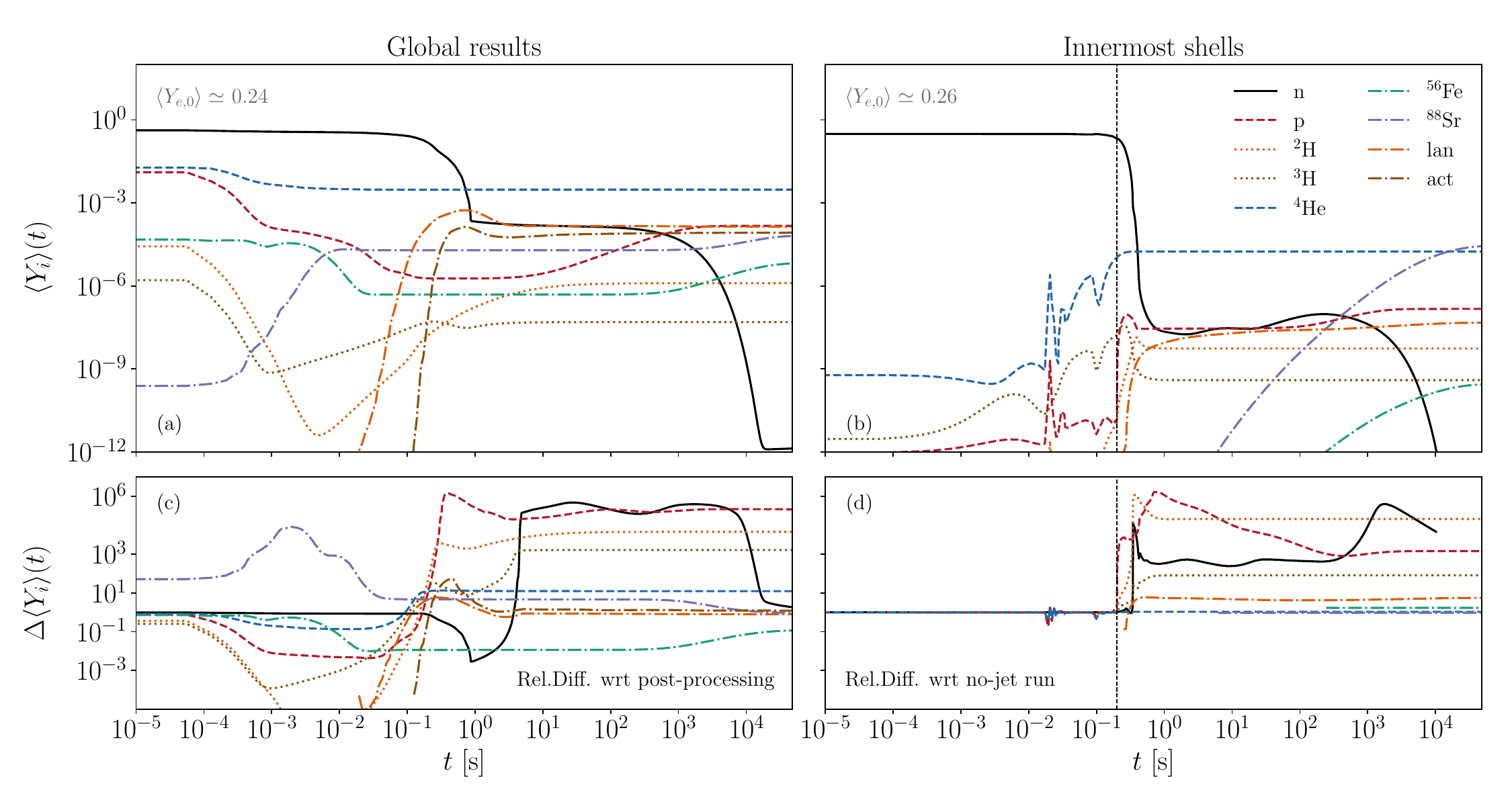}
	\caption{Evolution of the abundances of few selected isotopes and of the cumulative abundances of lanthanides and actinides. Panel ($a$) shows global results obtained with mass-weighted averages all over the ejecta. Panel ($b$) is restricted to an average over the innermost shell of all the angular sections in order to highlight the jet effects.
	Panel ($c$) shows the ratio $\Delta \langle Y_i \rangle = \langle Y_i \rangle / \langle Y_i^\text{pp} \rangle$ between our results and the prediction obtained with a post-process procedure similar to the one of \cite{Perego:2020evn}, see Appendix \ref{app:post-process} for details.
	Panel ($d$) shows a similar comparison between the results obtained with and without the jet.
		The vertical dashed lines in panels ($b$) and ($d$) indicate the time when the jet is launched.
	In panels ($c, d$), the results are only shown for times $\hat{t}$ such that $\langle Y_i \rangle (\hat{t}) > 10^{-12}$.
	} 							\label{fig:spec_abs}
\end{figure*}

In the bottom panel of Fig.~\ref{fig:final_abs} we show the final value of the relative difference
\begin{equation}
	\label{eq:rel_diff_abs}
	\Delta \langle Y_A^{\text{ch}} \rangle = \frac{\left| \langle Y_A \rangle - \langle Y_A^{\text{ch}} \rangle \right|}{\min{\left(\langle Y_A \rangle, \langle Y_A^{\text{ch}} \rangle \right)}}	
\end{equation}
\newtxt{between our results and the predictions $\langle Y_A^{\text{ch}} \rangle$} obtained with \newtxt{a \knec simulation of the ejecta with coupled NN and a prescribed homologous expansion. We evolve the system with the full hydrodynamic equations up to $t_0 = 10$~ms, and then force the density to evolve as $\rho(t,m) = \rho(t_0,m) (t/t_0)^{-3}$.}
\newtxt{Neglecting the details of the hydrodynamic evolution introduces an error of $\gtrsim 10\%$ for almost all values of mass numbers. The biggest discrepancies are observed for some light and second peak elements.}
\newtxt{We also compare our results against a post-process analysis similar to} the method described in \cite{Perego:2020evn}, \newtxt{but adapted to our initialization method.}
\newtxt{The agreement is overall worse in the new case, and the position of the third $r$-process peak is shifted.}
\newtxt{The extra source of error is introduced here by effectively coarsening the resolution on the initial thermodynamic conditions (see Appendix \ref{app:post-process} for details)}. 
Figure~\ref{fig:spec_abs} shows the evolution of the abundances for free protons and neutrons and a few selected isotopes. The $r$-process nucleosynthesis takes place between $t \sim 10^{-2} \textrm{ s}$ and $t \sim 1 \text{ s}$, when the first drop in $\langle Y_n \rangle$ occurs \newtxt{and the abundances of lanthanides and actinides saturate at $\langle Y_i \rangle \simeq 10^{-4}$}. The plateau at $Y_n \simeq 10^{-4}$ in panel ($a$) indicates the presence of regions of the ejecta where the material undergoes incomplete neutron burning. The second drop in the abundance of neutrons at $t \sim 10$~min, connected with an increase in the proton fraction, is due to free neutrons $\beta$-decays. 
\newtxt{Our analysis} shows qualitative features similar to those discussed in \cite{Perego:2020evn}.
However, a direct comparison between the two models shows discrepancies \newtxt{of several} orders of magnitude in the abundance evolution of various isotopes, see panel ($c$). \newtxt{In particular, note the six orders of magnitude difference in the neutron abundance for the post-nucleosynthesis plateau, and the overproduction of $^{56}$Fe from the post-processing of about two and one orders of magnitude at timescales between $10^{-2} - 10^3$~s and $t \sim 10^4$~s, respectively. At late times, the post-processing has underproduced $^4$He, $^2$H and $^3$H, and free protons of about 1, 3, and 6 orders of magnitude.}

The initial global abundance of $^4$He is of the order of $10^{-2}$ because of shells undergoing $\alpha$-rich freeze out\newtxt{, see panel ($a$)}.
This isotope is partially consumed at early times to construct heavier elements, \eg~$^{88}$Sr.
In the innermost shells, it is produced again during the $r$-process nucleosynthesis by $\alpha$-decays of the freshly produced heavy, neutron-rich elements, see panel ($b$).
These regions are crossed by the jet, which produces a rapid increase in the abundances of protons and deuterium at $t \simeq 200 \textrm{ ms}$. %
$^{88}$Sr is initially strongly suppressed, but it rises enough at early times to effectively act as a seed for $r$-process nucleosynthesis, see panel ($a$). At late times this element is produced again as part of the first peak by $\beta$-decaying neutron-rich isotopes. This is particularly evident in panel ($b$). We also show the abundance of $^{56}$Fe as representative of the most bounded nuclei around the iron peak. Its initial formation is favored by its high binding energy. Like strontium, $^{56}$Fe acts as a seed nucleus, but it is only partially produced again by later nuclear reactions.

The additional energy released by the jet significantly affects only a fraction of the inner shells at the highest latitudes. The mass involved is a too small fraction of the ejecta to yield visible effects on the global results presented in Fig.~\ref{fig:final_abs} and~\ref{fig:spec_abs}. In particular, the final nuclear yields are compatible to few percent between the complete simulations with and without the jet (see the bottom panel of Fig.~\ref{fig:final_abs}).
Nevertheless, the examination of the nucleosynthesis within the inner shells reveals some modifications of the final abundances for the selected elements of Fig.~\ref{fig:spec_abs}, see panel ($d$).
The sudden acceleration induced by the jet slows neutron captures down, therefore leaving, at the end of the nucleosynthesis, fewer lanthanides, more free $\beta$-decaying neutrons and, consequently, free protons. The faster dynamics also inhibits the burning of light elements, leading to an increase in the abundance of hydrogen isotopes at the end of the simulation.

\newpage
\subsection{Light curves}
\label{subsec:light_curves}

\begin{figure*}[t]
	\centering 
	\includegraphics[width=0.98\textwidth]{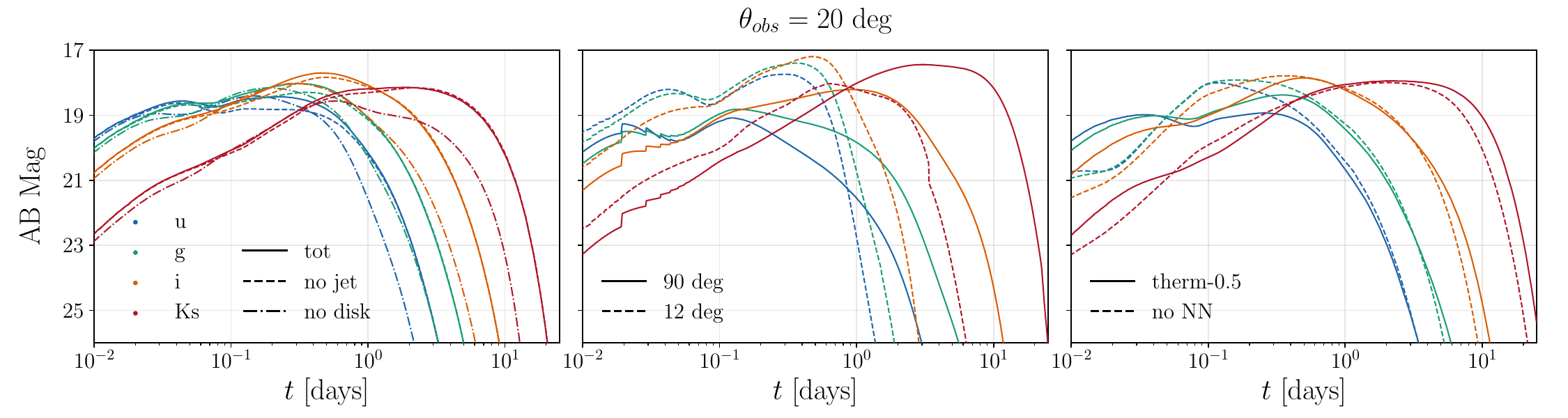}
	\caption{AB apparent magnitudes predicted by our simulation for the Gemini bands $u$, $g$, $i$ and $K_s$ for an observation angle of $\theta = 20$ degrees and a distance of $40$~Mpc.
	Left panel: results of the complete simulation (solid lines) against the predictions obtained by excluding the jet (dashed lines) or the disk component (dash-dotted lines).
	Central panel: isotropized $\theta = 12, 90$ degrees angular sections (solid and dashed lines respectively).
	Right panel: light curves obtained assuming a constant thermalization factor $f_\text{th} = 0.5$ with (solid lines) or without (with the same heating rate fits of \cite{Wu:2021ibi}, dashed lines) a coupled NN.}
	\label{fig:light_curves}
\end{figure*}

In Fig.~\ref{fig:light_curves} we show the kilonova light curves predicted by our model for a few selected UV/visible/IR bands for an observation angle of $\theta = 20$ degrees \newtxt{and a distance of $40$~Mpc}.

The disk is predicted to contribute to the blue component of the kilonova due to its lower opacity, see \eg~\citep{Metzger:2014ila, Siegel:2019mlp, Curtis:2023zfo}. However, at all angles, the photosphere enters the disk region only at late times ($t \sim 10$ days), when the temperature drop has already made the disk spectra red. Essentially, the  dynamical ejecta acts as a curtain, preventing the disk from significantly boosting the blue part of the kilonova.
Correspondingly, in the left panel of Fig.~\ref{fig:light_curves}, all the curves show an analogous decrease in their luminosity, when the disk is excluded, only due to the removal of part of the ejected mass.
The central panel of Fig.~\ref{fig:light_curves} shows that, around the equator, the opaque lanthanide curtain associated with the dynamical ejecta produces a red kilonova.
The blue components of the spectra become dominant at higher latitudes, suggesting a main contribution to the blue kilonova from the angular sections around the disk edge, where a still significative mass of relatively neutron-poor material is ejected.

\newtxt{To assess the impact of the online NN on the kilonova predictions, we plot in the right panel of Fig.~\ref{fig:light_curves} the results obtained without evolving the matter composition, but taking as input the heating rate fits employed already in \cite{Wu:2021ibi}}\footnote{
	\newtxt{Without the composition information given by the NN we cannot calculate the thermalization factors as described in Sec.~\ref{subsec:nnCoupling}. We therefore follow \cite{Wu:2021ibi} imposing a constant thermalization factor $f_\text{th} = 0.5$ and compare the results against a complete simulation with a coupled NN but with the same simple thermalization.}}.
Around $t \sim 4 \times 10^{-2}$ days, the simulation with a coupled NN shows a bump in the blue light curves, compatible with the UV precursor predicted in \cite{Metzger:2014yda, Combi:2022nhg} to be powered by $\beta$-decaying free neutrons.
\newtxt{This feature is not reproduced by the simulation with no coupled NN. The reason is that the fits of \cite{Wu:2021ibi} do not capture the free neutron contribution to the heating rate (see their Fig.~2 at $t \sim 10^{-2}$ days). Most importantly, the simulation without a coupled NN fails in predicting the position and brightness of the main peak of the high frequency filters. It predicts instead a faster dimming at late times of the low frequency ones. Thus, even if a coupled NN does have negligible impact on the hydrodynamic evolution, it introduces significative corrections in the predicted light curves.
While the heating rate competes with the pdV work and has thus a negligible effect on the hydrodynamics (see also Appendix \ref{app:NN_on_hydro}), it acts more directly on the luminosity of the optically thin regions above the photosphere. Moreover, relatively small changes in the temperature of the photosphere can lead to an observable shift on the frequency peak of its blackbody radiation.}

The jet slightly increases the early ($t \sim 0.1 - 1$ days) light curves in all bands. The effect is stronger and can be seen earlier for higher frequencies \newtxt{(see the left panel of Fig.~\ref{fig:light_curves})}. This is due to the extra energy input increasing the temperature and thus leading to a bluer and brighter emission.
A similar effect is discussed in \cite{Nativi:2020moj}, where the jet is found to clear a significant fraction of the lanthanide curtain.

\section{Conclusions}
\label{sec:conclusion}

In this work we investigated the nucleosynthesis process in a BNSM ejecta by means of a 2-dimensional, ray-by-ray radiation-hydrodynamics simulation incorporating an online NN.

Our results challenge the widely used approach of predicting $r$-process nucleosynthesis from isolated fluid elements.  Figure~\ref{fig:time_mass_comp} illustrates that it is not possible to fully predict the detailed distribution of the (final) nucleosynthesis yields solely from the initial thermodynamic conditions of a set of fluid elements. Moreover, the most commonly employed approximations for homologous expansions fail to capture the asymptotic hydrodynamic evolution of the unbound material. Therefore, even estimates for the nuclear heating rate obtained by post-processing isolated fluid elements, see \eg~\citep{Rosswog:2013kqa, Wanajo:2018wra, Wu:2021ibi, Rosswog:2022tus, Collins:2022ocl, Shingles:2023kua, Ricigliano:2023svx}, may lead to inconsistent results. To get a consistent picture of the ongoing nuclear processes, the effects related to radiation transfer and ejecta self-interaction must be taken into account.
\newtxt{On the other hand, the corrections introduced by an online NN on the nuclear energy supplied or removed to the expanding material do not introduce significative changes in the hydrodynamic evolution of the ejecta.
However, nuclear reactions have a more direct impact on their EM emission (see Fig.~\ref{fig:light_curves}).}
Providing a radiation-hydrodynamic simulation of the ejecta from compact binary mergers with an online NN thus appears crucial also for improving the realism of the \newtxt{predicted kilonova light curves}.

Some of the outer layers of the ejecta primarily consist of $\beta$-decaying free neutrons (see in Fig.~\ref{fig:time_mass_comp}), in agreement with \cite{Mendoza-Temis:2014mja, Radice:2018ghv}. In our simulations, having a very rapidly ($\tau \lesssim 5$~ms) expanding ejecta is not sufficient to prevent the production of third-peak elements. Very high entropies, $s \gtrsim 100 \, k_\text{B} / \text{baryon}$, are also required to leave a strong abundance of free neutrons after the $r$-process freeze-out.

From a qualitative point of view, our averaged results for nucleosynthesis yields in Fig.~\ref{fig:final_abs} align with other predictions in the literature, \eg~\citep{Cowan:2019pkx, Perego:2021dpw}. Our analysis reveals a connection between the production of light \newtxt{(below second peak)} and third-peak elements with late disk and dynamical ejecta respectively, in agreement with \cite{Martin:2015hxa, Cowan:2019pkx}.
Figure~\ref{fig:light_curves} shows that the outer lanthanides curtain shields the radiation produced by the inner disk ejecta, thus preventing them to give the significant contribution to the blue component of the kilonova predicted by \cite{Metzger:2014ila, Siegel:2019mlp, Curtis:2023zfo}.
The polar part of our ejecta evolves from higher $Y_e$ and produces few heavy $r$-process elements; this is consistent with the link discussed in \cite{Kasen:2017sxr, Tanaka:2017qxj, Nicholl:2017ahq, Perego:2017wtu} between the blue kilonova component and these sections of the unbound material.

\newtxt{We compared the final results for the nucleosynthesis yields with a \snec run where a homologous expansion is prescribed from $t = 10$~ms. Neglecting the details of the radiation-hydrodynamics evolution of the ejecta introduces} deviations $\gtrsim 10\%$ for all \newtxt{the mass numbers} (see the bottom panel of Fig.~\ref{fig:final_abs}).
\newtxt{In particular, the final abundances of some light ($A \lesssim 50$) and second $r$-process peak elements differ of almost one order of magnitude.}
We \newtxt{also} compared the time evolution of the nuclear abundances with a post-processing method analogous to a previous work that utilized the same NN \citep{Perego:2020evn}. We found significant quantitative differences for some of the analyzed elements (a qualitative agreement is found only for some specific shells in our simulation).
In particular, we found discrepancies of
\newtxt{several orders of magnitude in the evolution of $^4$He, $^{56}$Fe, deuterium, tritium, and free protons, see} Fig.~\ref{fig:spec_abs}. The differences are due \newtxt{both} to the coupling of the NN with the radiation-hydrodynamic evolution \newtxt{and to the grid introduced by the post-processing method on the initial thermodynamic conditions}.

The inclusion of an extra energy term mimicking a jet only affects the dynamic evolution of the polar regions of the ejecta, without altering the global, qualitative predictions of the $r$-process nucleosynthesis.

Our work will be extended and improved in several directions. We aim at implementing an improved thermalization of charged particles and a more realistic opacity treatment based on the complete, tracked information about matter composition. A following paper will report a systematic investigation of kilonova light curves using hundreds-of-milliseconds long numerical-relativity profiles and those improved thermalization and opacity models. Finally, the methods developed here will be ported to 3-dimensional simulations.

\begin{acknowledgments}
    FM acknowledges support from the Deutsche Forschungsgemeinschaft
  (DFG) under Grant No. 406116891 within the Research Training Group
  RTG 2522/1.
  FM also acknowledges ECT* for organizing the MICRA2023 workshop and
  TIFPA for their invitation to Trento in December 2023. Both occasions
  were sites of useful discussions.
  SB acknowledges funding from the EU Horizon under ERC Consolidator Grant,
  no. InspiReM-101043372 and from the Deutsche Forschungsgemeinschaft, DFG,
  project MEMI number BE 6301/2-1.
The work of AP is partially funded by the European Union
under NextGenerationEU, PRIN 2022 Prot. n. 2022KX2Z3B.
  DR acknowledges funding from the U.S. Department of Energy, Office of 
  Science, Division of Nuclear Physics under Award Number(s) 
  DE-SC0021177 and DE-SC0024388, and from the National Science 
  Foundation under Grants No. PHY-2011725, PHY-2020275, PHY-2116686, 
  and AST-2108467.

  Simulations were performed on the ARA and DRACO clusters at Friedrich Schiller
  University Jena, on the supercomputer SuperMUC-NG at the Leibniz-
  Rechenzentrum (LRZ, \url{www.lrz.de}) Munich, and on the national HPE
  Apollo Hawk at the High Performance Computing Center Stuttgart (HLRS).
  The ARA cluster is funded in part by DFG grants INST 275/334-1
  FUGG and INST 275/363-1 FUGG, and ERC Starting
  Grant, grant agreement no. BinGraSp-714626. 
  The authors acknowledge the Gauss Centre for Supercomputing
  e.V. (\url{www.gauss-centre.eu}) for funding this project by providing
  computing time on the GCS Supercomputer SuperMUC-NG at LRZ
  (allocations {\tt pn36ge} and {\tt pn36jo}).
  The authors acknowledge HLRS for funding this project by pro-
  viding access to the supercomputer HPE Apollo Hawk
  under the grant number INTRHYGUE/44215.
\end{acknowledgments}

\appendix
\twocolumngrid

\section{Hydrodynamic effects on NN}
\label{app:hydro_on_NN}

\begin{figure}	\centering 
	\includegraphics[width=0.48\textwidth]{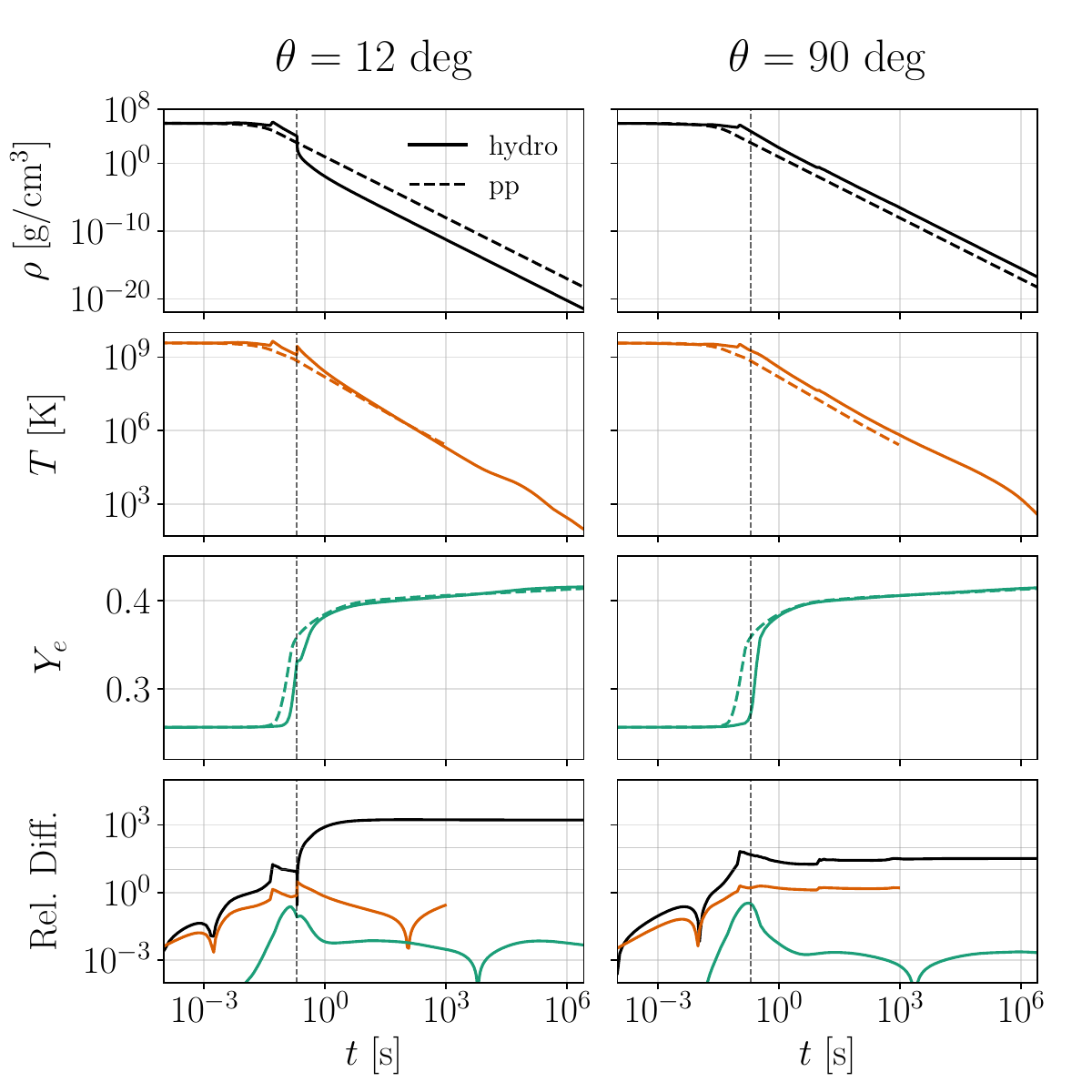}
	\caption{Evolution of the density (black), temperature (orange) and electron fraction (green) in the innermost shell of the $12$ degrees (left column) and $90$ degrees (right column) angular sections of the ejecta as calculated from the complete simulation (solid lines) and an independent \skynet run (dashed line). The bottom panels show the relative differences calculated with an expression analogous to Eq.~\eqref{eq:rel_diff_abs}. The evolution of the temperature from \skynet is plotted until it is actively updated by the NN and used in the nuclear calculations. The vertical dashed line at $t = 200$~ms indicates the launch of the jet.}
	\label{fig:single_fluidEl_evolution}
\end{figure}

To get an idea of how the prescribed thermodynamic trajectory and the hydrodynamical one differ, we compare in Fig.~\ref{fig:single_fluidEl_evolution} the density $\rho$, temperature $T$ and electron fraction $Y_e$ evolutions for the innermost shell of the $12, 90$ degrees angular sections of our \knec profile.
We find orders of magnitude discrepancies in $\rho$ and $T$, while $Y_e$ is generally more compatible between the two evolutions. Note, in the left column, how the jet affects the more polar trajectory, reducing the density and causing a discontinuity in the derivatives of $T (t)$ and $Y_e (t)$.
Mind that Fig.~\ref{fig:single_fluidEl_evolution} is only meant to work as an example, but is not representative of the general behavior of the ejecta. For instance, note in Fig.~\ref{fig:time_mass_comp} how $h$ settles at different final values (\ie~the expansion becomes homologous) at different times for different fluid elements. Better o worse agreement between the compared evolutions can be found looking at other mass shells and angular sections.

\section{NN effects on Hydrodynamics}
\label{app:NN_on_hydro}

\begin{figure}	\centering 
	\includegraphics[width=0.48\textwidth]{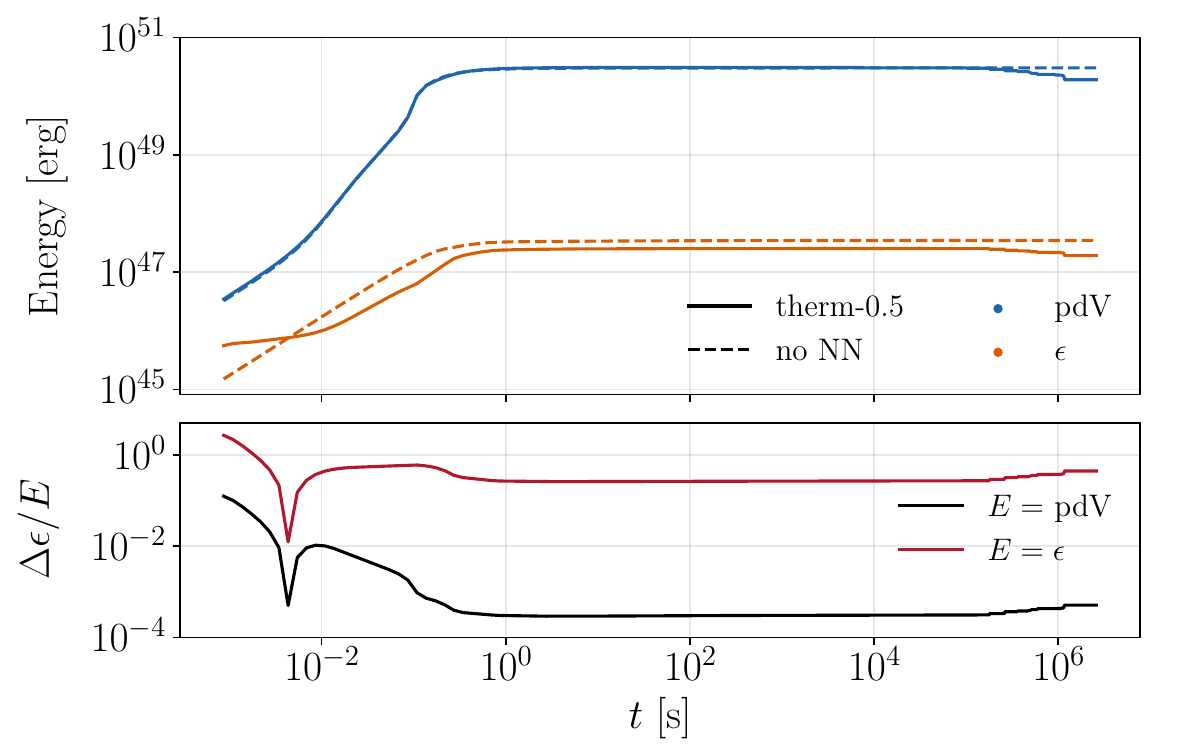}
	\caption{Top panel: evolution of total pressure work pdV (in blue) and nuclear heating $\epsilon = \dot{\epsilon}_\text{nucl} \, dt$ (orange) averaged all over the ejecta and integrated over time from the \knec runs with (solid lines) and without (dashed lines) a coupled NN and a constant thermalization factor $f_\text{th} = 0.5$.
	Bottom panel: difference in the nuclear heating between the two simulations normalized over pdV (black line) or $\epsilon$ (red line) from the simulation using the heating rate fits.}
	\label{fig:NN_on_hydro}
\end{figure}

We discuss here why the introduction of an online NN in a hydrodynamic simulation of the ejecta can affect the kilonova predictions while having negligible effects on the hydrodynamic evolution.
To do that, we run a \knec simulation using the heating rate fits of \cite{Wu:2021ibi} instead of the information from a coupled NN.
In Fig.~\ref{fig:NN_on_hydro}, we compare the total pressure work done on the ejecta and released nuclear energy, as a function of time, against a \knec run with an online NN. In both simulations we assume a constant thermalization factor $f_\text{th}=0.5$. The same assumption is made in \cite{Wu:2021ibi}, as the composition information needed by the detailed thermalization and coming from the NN are missing.

As discussed in Sec.~\ref{subsec:comp_evol}, the hydrodynamical evolution of the density profile from our complete run does not agree with the considered common prescription for homologous expansion. The same conclusion holds if the run using the heating rate fits is used.
The introduction of a coupled NN introduces global corrections in the heating rates of $\sim 10\%$. If specific fluid elements are considered, these corrections can grow up to a factor $10$. Anyway, as shown in Fig.~\ref{fig:NN_on_hydro}, the global corrections are typically $\mathcal{O}(10^{-4})$ the total pressure work done by the mass shells during their expansion. As a consequence, the effect of the online NN on their thermodynamic trajectories through Eq.~\eqref{eq:energy_cons} is negligible.
However, the kilonova light curves are affected by the NN, as the heating rate changes act more directly on the properties of the photosphere and on the radiation emitted by the shells outside of it (see Sec.~\ref{subsec:light_curves}).

\section{Post-processing} \label{app:post-process}

\cite{Perego:2020evn} compute the nucleosynthesis in post-processing using our same NN and imposing homologously expanding ejecta profiles.
A grid on the initial entropy $s_0$, expansion timescale $\tau_0$ and electron fraction $Y_{e,0}$ is introduced to have a series of ready-to-use tabulated results. 
To adapt their post-process procedure to our initialization method, we first distribute the mass shells of the initial ejecta profiles (the same we run with \knec\!\!, averaged over the azimuthal angle) on a grid based on the initial thermodynamic conditions.
We take the same grid in $s_0$, $\tau_0$ and $Y_{e,0}$ used in \cite{Perego:2020evn} and introduce a fine grid (150 values) on the temperature $T_0$ to be consistent with our NN initialization procedure.
We run then an independent instance of \skynet on each of the grid points. To be consistent with the homology assumption of \cite{Radice:2016dwd}, we base our grid on the value $\tau_0 = R_E / v_0$, with $R_E$ extraction radius and $v_0$ initial velocity, initialize each NN at $t_0=\tau_0$ and $T=T_0$ imposing NSE, and evolve it with the prescription $\rho(t) = \rho_0 ((t+\tau_0)/\tau_0)^{-3}$.

The introduction of the grid on $s_0$, $\tau_0$, $Y_{e,0}$ coarsens the resolution on the initial thermodynamic conditions, thus introducing another layer of approximation in the post-processing on top of the assumption of homologous expansion. 		As discussed in Sec.~\ref{subsec:nucl_yields}, this is also a relevant source of error (note anyway that a step that reduces the number of NN to run is crucial to define a post-process procedure able to give results in a reasonable amount of time).

\end{document}